\documentclass{article}
\usepackage{spconf,amsmath,graphicx}
\usepackage{booktabs}
\usepackage{enumitem}
\setlist{nosep, leftmargin=14pt}

\usepackage{mwe} 

\title{Pediatric TSC-Related Epilepsy Classification from Clinical MR Images Using Quantum Neural Network}
\name{
	\parbox{\linewidth}{\centering
	Ling Lin$^{1,2}$, Yihang Zhou$^{2,3}$, Zhanqi Hu$^{4}$, Dian Jiang$^{3}$, Congcong Liu$^{1,2}$, Shuo Zhou$^{1,2}$,\\ Yanjie Zhu$^{1,2}$, 
	Jianxiang Liao$^{4}$, Dong Liang$^{1,2,3}$, Hairong Zheng$^{1,2}$, Haifeng Wang$^{1,2}$\thanks{Ling Lin, Yihang Zhou, and Zhanqi Hu contributed equally to this study and are the co-first authors. The corresponding author is Haifeng Wang (hf.wang1@siat.ac.cn).}}}
\address{
	$^{1}$Paul C. Lauterbur Research Center for Biomedical Imaging, Shenzhen Institute of \\ Advanced Technology, Chinese Academy of Sciences, Shenzhen, Guangdong, China \\
	$^{2}$University of Chinese Academy of Sciences, Beijing, China \\
	$^{3}$Center for Medical AI, Shenzhen Institute of Advanced Technology, Chinese Academy of\\ Sciences, Shenzhen, Guangdong, China\\
	$^{4}$Department of Neurology, Shenzhen Children's Hospital, Shenzhen, Guangdong, China
}

\begin{document}
%
\maketitle
\begin{abstract}
Tuberous sclerosis complex (TSC) manifests as a multisystem disorder with significant neurological implications. This study addresses the critical need for robust classification models tailored to TSC in pediatric patients, introducing QResNet,a novel deep learning model seamlessly integrating conventional convolutional neural networks with quantum neural networks. The model incorporates a two-layer quantum layer (QL), comprising ZZFeatureMap and Ansatz layers, strategically designed for processing classical data within a quantum framework. A comprehensive evaluation, demonstrates the superior performance of QResNet in TSC MRI image classification compared to conventional 3D-ResNet models. These compelling findings underscore the potential of quantum computing to revolutionize medical imaging and diagnostics.Remarkably, this method surpasses conventional CNNs in accuracy and Area Under the Curve (AUC) metrics with the current dataset. Future research endeavors may focus on exploring the scalability and practical implementation of quantum algorithms in real-world medical imaging scenarios. 
\end{abstract}
\begin{keywords}
Quantum Computing, CNN, Tuberous sclerosis complex, MRI, Image Classification
\end{keywords}
\section{Introduction}
\label{sec:intro}

Tuberous sclerosis complex (TSC) is an inherited multisystem disorder characterized by cellular and tissue dysplasia in various organs\cite{shen2022nerp}. The underlying pathogenesis involves hyperactivation of the mTOR pathway, stemming from de novo or inherited mutations in the TSC1 or TSC2 genes\cite{uysal2020tuberous}. Approximately 50\% of TSC patients exhibit developmental delay or intellectual disability\cite{krueger2013tuberous}. Additionally, TSC-Associated Neuropsychiatric Disorders (TANDs), which encompass behavioral, psychiatric, neuropsychological, and social/emotional processing issues, affect around 90\% of patients\cite{de2018tsc}. Nearly half of the patients ($\sim$50\%) receive a diagnosis of autism spectrum disorder, and almost two-thirds experience their initial seizure in the first year of life\cite{yang2020machine}. These complexities underline the urgent need for efficient, reliable classification models for TSC, particularly in pediatric cases.

Structural brain abnormalities are prevalent in TSC patients, manifesting as cortical or subcortical tubers, subependymal nodules (SENs), subependymal giant cell astrocytomas (SEGAs), and white matter radial migration lines (RMLs)\cite{feliciano2020neurodevelopmental}.Magnetic resonance imaging (MRI)  is a modality used routinely to diagnose TSC. Tubers, detectable in approximately 80–90\% of TSC patients in MRI, constitute a major diagnostic criterion\cite{lu2018central}. These malformations appear as focal regions on MRI with categorically distinct changes in T1, T2, and fluid-attenuated inversion recovery (FLAIR) sequences, most prominently hypointense in T1 and hyperintense in T2 weighted images\cite{feliciano2020neurodevelopmental}. 
\begin{figure*}[t]
	\centering
	\includegraphics[width=0.75\textwidth, keepaspectratio=true]{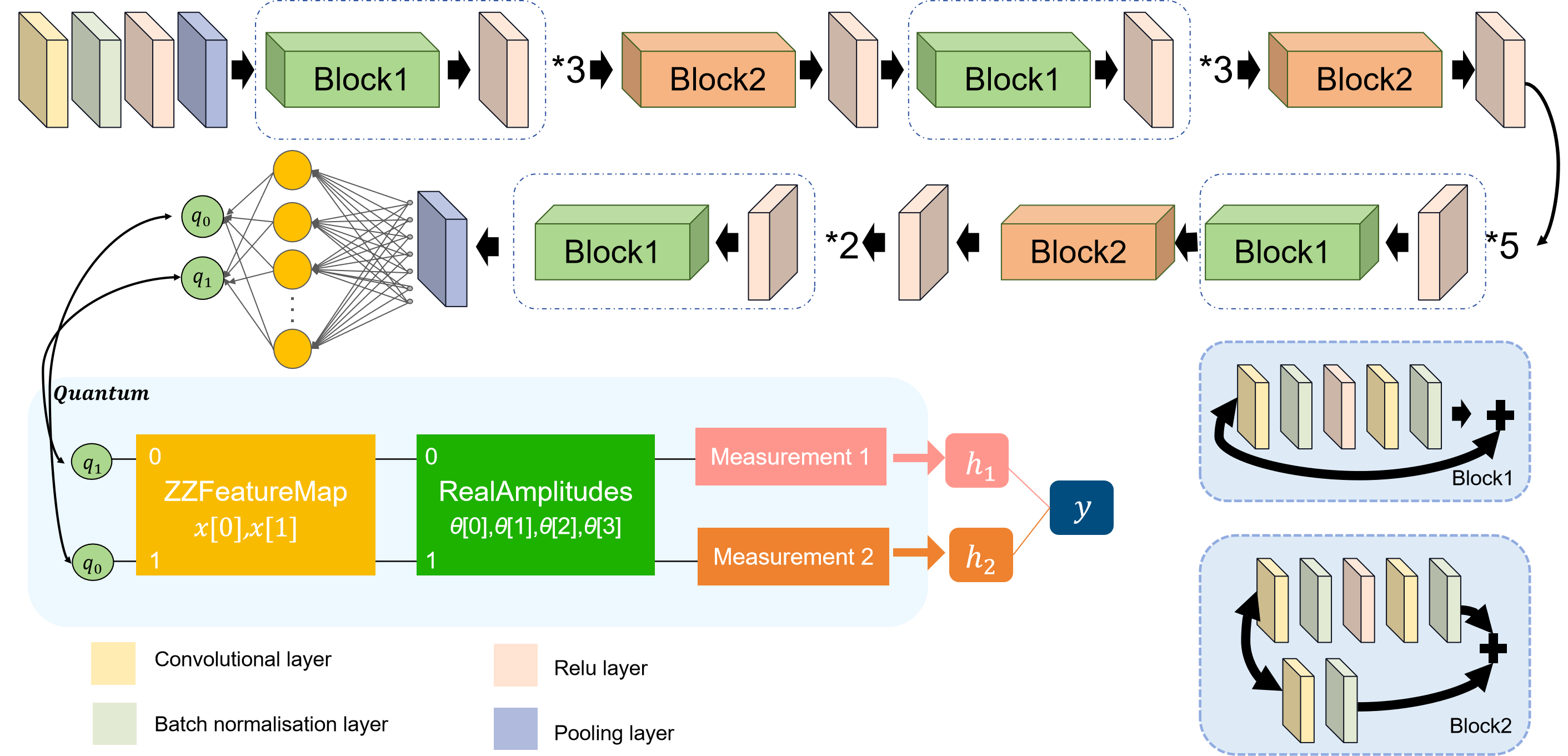}
	\caption{The Layers Structure of The Proposed QResNet}
	\label{figure1}
\end{figure*}
While prior studies have demonstrated the success of deep convolutional neural network (CNN) models in accurately classifying lung cancer\cite{lu2018central}, bone lesions\cite{eweje2021deep}, and predicting brain age on MRI, the limited availability of images presents a challenge in rare disorders such as TSC. The existing literature lacks comprehensive studies utilizing CNNs for identifying TSC patients. With the growing complexity of data and the need for advanced models, researchers are exploring alternative approaches, including quantum computing. Recent studies aim to improve methods for identifying TSC patients through imaging techniques, recognizing the limitations of CNN models on smaller datasets. Quantum deep learning, the challenge of creating quantum circuits to enhance neural network operations, has been explored in various works. Quantum computing, which leverages principles from quantum mechanics such as entanglement, superposition, and interference, offers faster and more powerful processing capabilities\cite{tantawi2023exploring}. Quantum machine learning techniques, including Quantum Neural Networks, have emerged, holding the potential to revolutionize medical imaging and diagnostics. In our research, we introduce a novel deep learning model called QResNet, seamlessly integrating conventional convolutional neural networks with quantum neural networks. This model aims to discern TSC patients through the analysis of MRI data.

\section{Method}

Different variants of CNNs has exhibited remarkable classification performance in image classification. The traditional 2D convolution model requires slicing MRI and cannot make good use of the spatial characteristics of MRI. In this paper,3D Convolutional neural networks and the quantum layer (QL) are combined to identify TSC patients.

In the quantum layer, a two-layer architecture is utilized. The two fundamental quantum layers that make up this two-layer design are the ZZFeatureMap and the Ansatz layers. These layers work together to process and analyze classical data while taking advantage of the unique features of quantum computing. In traditional machine learning, feature maps are used to transform raw input data into a higher dimensional feature space. Similarly, quantum feature maps are used to transform classical data into a quantum state that can be processed by a quantum computer. In a quantum feature map, the input data is transformed using a quantum gates operation to produce a new quantum state vector that contains higher-order correlations between the original data points. The quantum feature maps are able to efficiently generate complex transformation that are computationally hard to construct using classical method. Moreover, the base quantum circuit operation can also be repeated multiple times to construct more complex feature maps. ZZfeatureMap is the most common proposed feature maps because of its simplicity and experimental performance. The ZZFeatureMap can be mathematically represented as follows:
\begin{align}
	U(\theta) = \prod_{k} U_{k}(\theta_{k}) \label{eq1}
\end{align}

where $\theta_{k} = x_{i} (2\pi - x_{j}) \quad \text{for } i<j $, and $U_{k}(\theta_{k}) $ is a two-qubit ZZgate.

The resultant quantum state encapsulates the essential features of the incoming data, priming it for subsequent processing by the ensuing Ansatz layer. The Ansatz layer, also denoted as the variational layer, undertakes the processing of the quantum state generated by the ZZFeatureMap. In the QResNet architecture, the RealAmplitudes circuit is employed as the Ansatz. This circuit comprises single-qubit Y-rotations and entangled CNOT gates. Through the analysis of Y-rotations, which are governed by programmable parameters dynamically adjusted during training, the QResNet can adapt its behavior by learning from the input data. The entangling gates facilitate the connection of qubits, fostering the interactions requisite for the intricate quantum behavior essential for quantum computation. In mathematical terms, the RealAmplitudes circuit is expressed as follows:
\begin{align}
	\nonumber
	V(\theta) = RZ(\Phi[2n-1],n) CX(n,n-1) \\ 
	RY(\Phi[2n],n-1) CX(n,n-1) \label{eq2}
\end{align}

where RZ and RY are single-qubit rotation gates, CX is the two-qubit CNOT gate, $\Phi$ is a
parameter vector, and n ranges from 1 to the number of qubits in the circuit.
The two-layer QL model's ZZFeatureMap (U) and Ansatz layer (V) provide a powerful way to analyse and learn from classical data in a quantum setting.
The proposed QResNet's layer structure is displayed in Fig.1. Convolutional, batch normalisation, and fully connected layers are among the layers that make up conventional convolutional neural networks. The final fully connected layer, whose output is the class, serves as the quantum layer's input.
The Qiskit Python library is used to execute the suggested QResNet on the Qiskit Aer simulator. This simulator allows for the simulation of quantum circuits and the execution of quantum algorithms on conventional hardware. The proposed QResNet circuit is implemented and simulated using the Qiskit Aer simulator by generating convolutional layers, pooling layers, and other procedures. Before attempting to implement quantum machine learning techniques on actual quantum hardware, this simulator can be a useful tool for testing them.

\section{Data Description}

This investigation enrolled a cohort exclusively sourced from Shenzhen Children’s Hospital spanning the period from January 2013 to January 2021. The study encompassed a total of 520 pediatric participants, comprising 260 individuals diagnosed with tuberous sclerosis complex (TSC) and an equivalent number of 260 healthy children (HC).


 All subjects provided written informed consent prior to participating, and the study protocols received approval from the Institutional Review Board (IRB) at Shenzhen Institutes of Advanced Technology, Chinese Academy of Sciences. The imaging dataset encompassed fluid-attenuated inversion recovery (FLAIR) images and T2-weighted images (T2w).

\begin{figure}[t]
	\centering
	\includegraphics[scale=0.25]{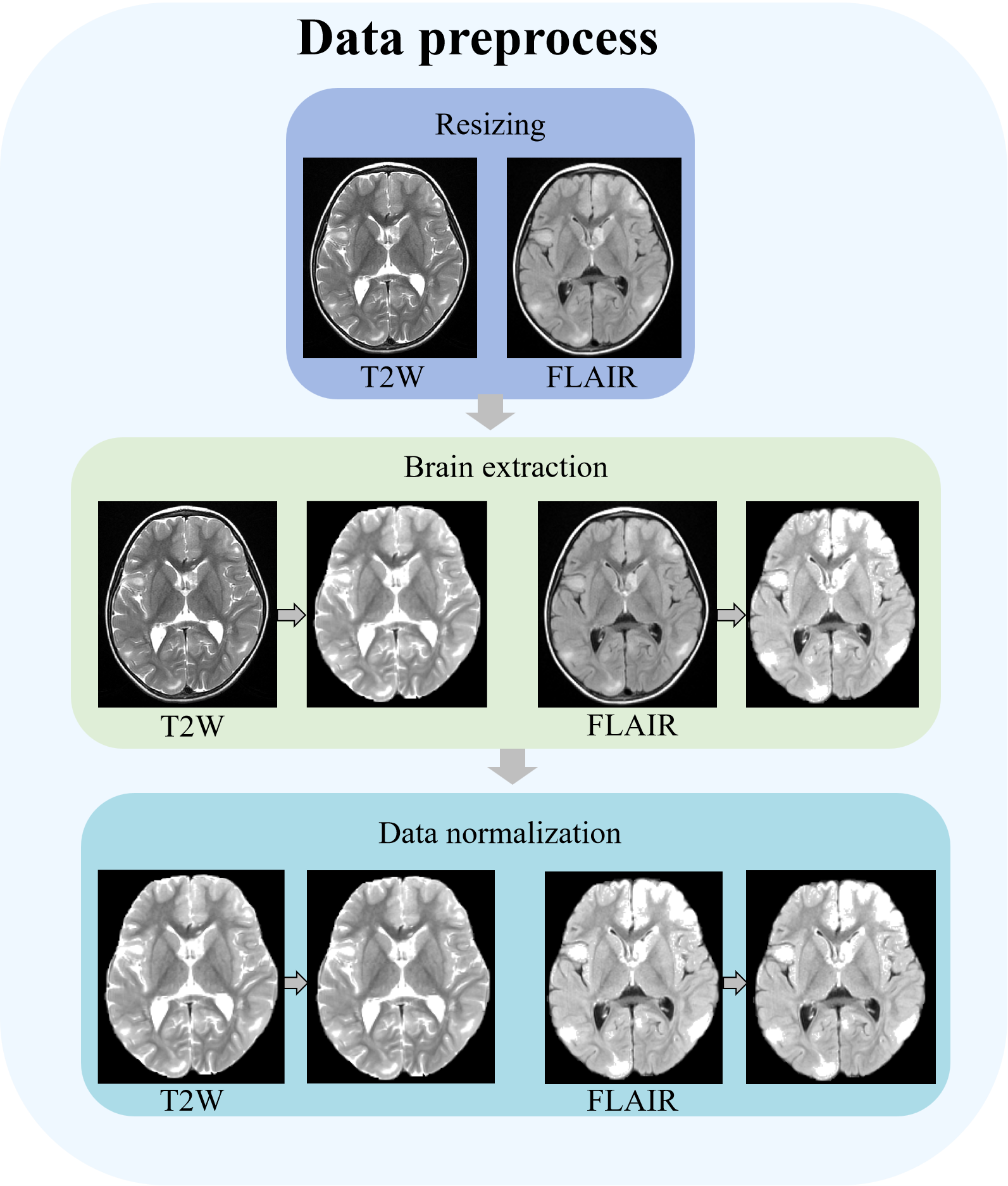}
	\caption{Schematic of the data preprocessing pipeline}
	\label{figure3}
\end{figure}

To facilitate robust model training and evaluation, the datasets were systematically partitioned into training (65\%, n = 334), validation (15\%, n = 82), and test (20\%, n = 104) subsets. Stratified random sampling was employed to maintain consistent class ratios across all sets, ensuring representative distribution. Importantly, there was no overlap of patients between the training, validation, and test sets, preserving the independence of each dataset partition.
The main patient characteristics of all 520 cases are listed in Table 1. Of the 260 TSC patients, 148(56.9\%) were male, and mean age is 3.87. Of the 260 HC, 144(55.4\%) were male, and mean age is 6.43.

\begin{table}[h]
	\centering
	\small
	\begin{tabular}{cccc}
		\toprule
		\textbf{} & \textbf{TSC} & \textbf{HC} & \textbf{P-value} \\ 
		\midrule
		No. & 260 & 260 &   \\ 
		Male, n(\%) & 148(56.9\%) & 144(55.4\%) & 0.834 \\
		Age,mean$\pm$SD(years) & 3.87 $\pm$ 3.63 & 6.43 $\pm$ 4.02 & $<0.001$ \\ 
		\bottomrule
		
	\end{tabular}
	\caption{Demographic characteristics of the 520 subjects}
\end{table}

we first strip the skull in MRI using a deep-learning tool HD-bet\cite{isensee2019automated}, which may be helpful for classification tasks. The size of all 3D MRI images was resized to (128,128,128), and the image intensity was then normalized to the range between 0 and 1. Fig.2 shows the schematic of the data preprocessing pipeline.


\section{Result}
\label{sec:typestyle}

In our study, we evaluated the classification performance of each model across different cohorts using a set of key metrics: Area Under the Curve (AUC), accuracy (ACC), sensitivity (SEN), and specificity (SPE). 
This segment highlights the effectiveness of the novel QResNet model in classifying TSC MRI images. A comprehensive assessment framework was employed to compare the performance of QResNet with the conventional 3D-ResNet34 model. Table I provides a detailed comparison of these two models, demonstrating their efficacy in TSC MRI image classification.
\begin{table}[h]
	\centering
	\begin{tabular}{cccccc}
		\toprule
		\textbf{Input} & \textbf{Model} & \textbf{AUC} & \textbf{ACC} & \textbf{SEN} & \textbf{SPE} \\ 
		\midrule
		FLAIR & ResNet3D &  0.982 & 0.923 & 0.942 & 0.904 \\ 
		T2W & ResNet3D & 0.977 & 0.894 & 0.923 & 0.865 \\
		FLAIR & QResNet & 0.995 & 0.962 & 0.942 & 0.981 \\
		T2W & QResNet & 0.984 & 0.933 & 0.904 & 0.962 \\
		\bottomrule
	\end{tabular}
	\caption{The classification performance of different models in test the set}
	\label{tab2}
\end{table}

The test results in Table I demonstrate the superior performance of the QResNet model. An extensive analysis was conducted by separately inputting Fluid-attenuated inversion recovery (FLAIR) and T2 weighted (T2w) images into the models. Notably, when analyzing FLAIR images, QResNet achieved impressive accuracy and AUC scores, recording an accuracy of 0.976 and an AUC of 0.995. These metrics significantly outperformed those of other models. Furthermore, with T2w images as input, QResNet maintained high performance, achieving an accuracy of 0.963 and an AUC of 0.983. The ROC curve and AUC for QResNet, illustrated in Fig.3, showcase its performance when provided with T2-weighted (T2W) and fluid-attenuated inversion recovery (FLAIR) images. In both FLAIR and T2W inputs, QResNet's performance exceeded that of the traditional 3D-ResNet model, highlighting its potential as an effective tool for TSC MRI image classification.

\begin{figure}[h]
	\centering
	\includegraphics[scale=0.25]{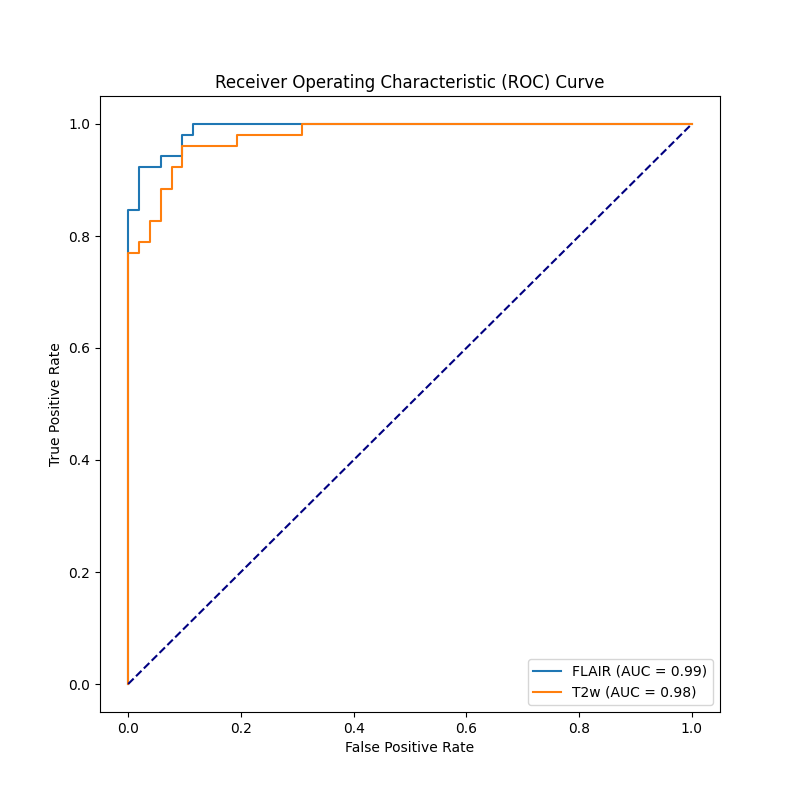}
	\caption{The ROC curves of the proposed QResNet}
	\label{figure4}
\end{figure}
\section{Discussion}
This study contributes to the field of medical imaging and TSC classification by introducing a novel QResNet approach. The integration of quantum computing principles into deep learning models holds promise for advancing medical diagnostics, particularly in the challenging domain of rare disorders like TSC. Future research directions may explore the scalability and practical implementation of quantum algorithms in real-world medical imaging scenarios.


\bibliography{strings}

\begin{thebibliography}{1}

\bibitem{uysal2020tuberous}
Sanem~Pinar Uysal and Mustafa {\c{S}}ahin,
\newblock ``Tuberous sclerosis: a review of the past, present, and future,''
\newblock {\em Turkish journal of medical sciences}, vol. 50, no. 10, pp.
  1665--1676, 2020.

\bibitem{krueger2013tuberous}
Darcy~A Krueger, Hope Northrup, Steven Roberds, Katie Smith, Julian Sampson,
  Bruce Korf, David~J Kwiatkowski, David Mowat, Mark Nellist, Sue Povey,
  et~al.,
\newblock ``Tuberous sclerosis complex surveillance and management:
  recommendations of the 2012 international tuberous sclerosis complex
  consensus conference,''
\newblock {\em Pediatric neurology}, vol. 49, no. 4, pp. 255--265, 2013.

\bibitem{de2018tsc}
Petrus~J De~Vries, Elena Belousova, Mirjana~P Benedik, Tom Carter, Vincent
  Cottin, Paolo Curatolo, Maria Dahlin, Lisa D’Amato, Guillaume~B
  d’Aug{\`e}res, Jos{\'e}~C Ferreira, et~al.,
\newblock ``Tsc-associated neuropsychiatric disorders (tand): findings from the
  tosca natural history study,''
\newblock {\em Orphanet journal of rare diseases}, vol. 13, pp. 1--13, 2018.

\bibitem{yang2020machine}
Jun Yang, Cailei Zhao, Shi Su, Dong Liang, Zhanqi Hu, Haifeng Wang, and
  Jianxiang Liao,
\newblock ``Machine learning in epilepsy drug treatment outcome prediction
  using multi-modality data in children with tuberous sclerosis complex,''
\newblock in {\em 2020 6th International Conference on Big Data and Information
  Analytics (BigDIA)}. IEEE, 2020, pp. 100--103.

\bibitem{feliciano2020neurodevelopmental}
David~M Feliciano,
\newblock ``The neurodevelopmental pathogenesis of tuberous sclerosis complex
  (tsc),''
\newblock {\em Frontiers in Neuroanatomy}, vol. 14, pp. 39, 2020.

\bibitem{lu2018central}
Derek~S Lu, Patrick~J Karas, Darcy~A Krueger, and Howard~L Weiner,
\newblock ``Central nervous system manifestations of tuberous sclerosis
  complex,''
\newblock in {\em American Journal of Medical Genetics Part C: Seminars in
  Medical Genetics}. Wiley Online Library, 2018, vol. 178, pp. 291--298.

\bibitem{eweje2021deep}
Feyisope~R Eweje, Bingting Bao, Jing Wu, Deepa Dalal, Wei-hua Liao, Yu~He,
  Yongheng Luo, Shaolei Lu, Paul Zhang, Xianjing Peng, et~al.,
\newblock ``Deep learning for classification of bone lesions on routine mri,''
\newblock {\em EBioMedicine}, vol. 68, 2021.

\bibitem{tantawi2023exploring}
Baraa Tantawi, Hamza~Kamel Ahmed, Malak Magdy, Mohamed Adel, and Gehad~Ismail
  Sayed,
\newblock ``Exploring the power of quantum convolutional neural networks for
  brain mri image classification,''
\newblock in {\em 2023 Intelligent Methods, Systems, and Applications (IMSA)}.
  IEEE, 2023, pp. 549--584.

\bibitem{isensee2019automated}
Fabian Isensee, Marianne Schell, Irada Pflueger, Gianluca Brugnara, David
  Bonekamp, Ulf Neuberger, Antje Wick, Heinz-Peter Schlemmer, Sabine Heiland,
  Wolfgang Wick, et~al.,
\newblock ``Automated brain extraction of multisequence mri using artificial
  neural networks,''
\newblock {\em Human brain mapping}, vol. 40, no. 17, pp. 4952--4964, 2019.

\end{thebibliography}
\bibliographystyle{IEEEbib}

\end{document}